\input aa.cmm
\voffset=0 truecm
\input psfig.tex
\overfullrule=0 pt
%\refereelayout
\MAINTITLE{Morphological classification and structural parameters 
of galaxies in the Coma and Perseus clusters\FOOTNOTE{
Based on observations made with the 2-meter T\'elescope Bernard Lyot of 
Pic-du-Midi Observatory, operated by INSU (CNRS) and the Schmidt
telescope at the Calern Observatory (OCA). All
tables and figures are only available in electronic form at the
CDS anonymous ftp to cdsarc.u-strasbg.fr (130.79.128.5) or via
http://cdsweb.u-strasbg.fr/Abstract.html}
}
\AUTHOR{S. Andreon@1@2, E. Davoust@2, P. Poulain@2}
\INSTITUTE{@1 Osservatorio Astronomico di Capodimonte, via Moiariello 16, 80131 Napoli, 
            Italy\FOOTNOTE{Present address of SA},
            andreon at cerere.na.astro.it\newline
           @2 CNRS-UMR 5572, Observatoire Midi-Pyr\'en\'ees, 14, Av. E.
            Belin, 31400 Toulouse, France, davoust, poulain at
obs-mip.fr\newline
                     }
\DATE{ Accepted, February, 2, 1997} 
 
\ABSTRACT{
We present the results of an isophotal shape analysis of galaxies in the
Coma and Perseus clusters.  These data, together with those of two
previous papers, provide two complete samples of galaxies with reliable
Hubble types in rich clusters: 

1) all galaxies brighter than m$_b$ = 16.5 falling within one degree (=2.3 
Mpc) from the center of the Coma cluster (187 galaxies), 

2) all galaxies brighter than m$_{Zwicky}=15.7$ in a region of $5\degr 3' 
\times 5\degr 27'$ around the center of the Perseus cluster (139 galaxies). 

These two complete samples cover 5 orders of magnitude in galaxy density and 
span areas of 91 and 17 Mpc$^2$, clustercentric radii up to 2.3 and 6.4 
Mpc, for Perseus and Coma respectively. They will be used in subsequent 
papers to study the dependence of galaxy types on cluster environment and as 
reference samples in comparisons with distant clusters. } 


\THESAURUS{20(11.03.4 (Abell 1656, Coma cluster), (Abell 426, Perseus
cluster), 11.05.1, 11.06.2)}

\maketitle 
\titlea {Introduction}

In two previous papers, we presented an isophotal shape analysis with 
morphological type estimates for a large number of galaxies in two clusters of 
galaxies.  Our analyses, based on CCD images and photographic plates, 
concerned more than 200 galaxies in three regions of Coma (Andreon et al. 
1996, hereafter Paper I) and about 100 galaxies in the Perseus cluster (Poulain,
Nieto \& Davoust 1992).  The aim of both works was to collect as many 
galaxies as possible in these two clusters for subsequent
studies of galaxy properties (e.g. Michard 1996; Andreon 1994, 1996; Andreon, 
Davoust \& Heim 1997). The galaxies observed with a CCD were selected mainly 
because they had previously been classified as early-type. 
 
Once the data accumulate and the samples grow to a reasonable size, it becomes 
possible to trace some notable properties of the Hubble types, but we 
convinced ourselves that slight differences among the properties of the 
morphological types were not properly measured with {\it incomplete} 
samples.  Using samples with various degrees of completeness, different 
authors have in fact reached different conclusions on many galaxy properties: 
on the galaxy mean surface brightness of the types (Andreon 1996), on the 
optical luminosity function of boxy Es and disky Es (Andreon 1994 and 1996), 
on the radio luminosity function of boE and diE (Lowen \& Owen 1995). The two 
previously available samples of galaxies in Coma and Perseus, although large 
in size and almost complete in (small) selected areas and within restricted 
magnitude ranges, were not quite complete in magnitude. Therefore we decided 
to observe the galaxies unobserved in these previous surveys. We also 
reobserved some galaxies whose detailed morphological type was unsatisfactory. 

Since it was beyond our observing capabilities to complete the two samples 
down to the magnitude of the faintest observed galaxy and out to the distance 
of the most peripheral galaxy, we set the more limited goal of completing our 
samples within an area and down to a magnitude limit that does not exclude too 
many faint galaxies already observed by our team and at the same time does not 
include an unreasonable number of new galaxies to be observed. 

In order to complete the observation of the samples rapidly, we gave up the 
idea of obtaining independent morphologies from images in different passbands 
(e.g. $V$ and $r$) and from different observing material (CCD images and 
small-scale plates), properties characterizing the two previous surveys. 
Furthermore, since high resolution conditions are not necessary for 
classifying spiral galaxies with obvious spirals arms, we also made use of 
Schmidt plates, thus innovating with respect to our two other papers. 

The availability of a thinned CCD at the 2-meter T\'elescope Bernard Lyot 
(hereafter TBL), with a good quantum efficiency in blue and the fact that 
spiral galaxies are easier to classify in visible 
than in red, prompted us 
to observe the program galaxies in Johnson $V$, instead of Gunn $r$. The 
increased CCD quantum efficiency largely compensates for the decrease in 
luminosity of early-type galaxies from $r$ to $V$, allowing us to image the 
program galaxies in $V$ within the allocated telescope time. 

The paper is organized as follow. We present the data completing the two
samples of galaxies in Coma and in Perseus in Sect. 2. The techniques of
analysis used in the present study are briefly summarized in Sect. 3. The
results are given in the form of tables (Tables 2, 3, 4 and 5) and
presented only in electronic form; the tables include global photometric
and geometrical parameters as well as detailed morphological information,
but not the photometric and geometrical profiles as a function of radius.
Notes on individual galaxies are included in the Tables.  An estimate of the
quality of our measures is presented in Sect. 4, and the results of the
paper are summarized in Sect. 5. 

We adopt a Hubble constant of $H_0=50$ km s$^{-1}$ Mpc$^{-1}$.

\titlea {The sample}

\titleb {Coma}

This sample of galaxies was taken from the catalogue of Godwin et al.
(1983;  hereafter GMP), which lists all galaxies down to magnitude m$_b$ =
20.0 in a 2.63 degree square area, centered on the Coma cluster.  We
selected all the galaxies down to magnitude m$_b$ = 16.5, within one
degree from the cluster center whose morphological type is not published
in Paper I.  Furthermore we reobserved all SAB0 galaxies (i.e. S0s for
which the presence of a bar is uncertain) and unE (i.e. Es where boxiness
or diskiness is undetermined) observed under mediocre seeing conditions by
Andreon et al.  (1996), as well as other galaxies of borderline type
between two classes. We discarded from the observations 3 galaxies (GMP
978, 1646, 1741) which do not belong to the Coma cluster, as they have
velocities\fonote{Velocities of the program galaxies in Coma and Perseus
were mainly collected from public databases, such as NED and LEDA, and by
a survey of the literature.} of more than 4000 km s$^{-1}$ relative to the
cluster center. Figure 1 (only available in electronic form) shows the spatial
distribution of the
Coma galaxies.  With the present sample, all (187) galaxies brighter than
$M_b=-19.2$ mag in the surveyed area have been observed. 

The galaxies of this sample were first inspected on two Schmidt plates, OCA 
\#2849 and OCA \#2842, taken at the Calern Observatory and kindly provided by 
C. Pollas. These plates were digitized at the MAMA\fonote{MAMA (Machine 
Automatique \`a Mesurer pour l'Astronomie) is operated by CNRS/INSU.} with a 5 
and 10 micron (= 0.33{\tt "} and 0.65{\tt "}) step respectively, and with 
an aperture of 10 microns. The digitization produces image files with pixel 
readings proportional to plate density. 

All galaxies with evident asymmetric or irregular isophotes or with spiral
arms were classified as S and eliminated from further observations with CCDs. 
We nevertheless already had CCD images for half of them, because they were 
published by other authors or in the field of view of other program galaxies. 
The CCD images confirm the type estimates based on our Schmidt plates for all 
galaxies; this confirms the reliability of this method for classifying obvious 
spiral galaxies, even at the distance of the Coma cluster. 

Finally, we found 5 galaxies in our CCD-image archives which had not been 
classified in Paper I; these 5 were not reobserved. 

Three suspected spiral galaxies were observed in December 1994 with the
TBL, and the remaining 28 galaxies were observed during one observing run
in February 1995 with the same instrument and setup. The CCD was a
Tektronics $1024\times1024$, with a pixel size of 24 $\mu$ corresponding
to 0.30 arcsec on the sky. The seeing was 1.0--1.1 arcsec (FWHM) and the
nights were photometric.  The exposure time was 20 minutes for all
galaxies. With the present observations, the median seeing of the complete
sample reduces from 1.48 arcsec in Paper I to 1.2 arcsec, which is a large
improvement since less than one third of the sample had been observed.
This corresponds to a restframe resolution of 0.75 Kpc, not very different
from the resolution that the {\it Hubble Space Telescope} offers for
distant ($z\sim0.4$) galaxies. 

\titleb {Perseus}

The sample is composed of all (141) galaxies listed in the Zwicky
catalogue (Zwicky 1961-1968) in a box of $5\degr 3' \times 5\degr 27'$
centered on $3^h14^m42^s$, $41\degr13'30"$ to which we added two Tiff
(1977) galaxies. NGC 1233 is listed twice in the Zwicky catalogue (Zw
525-6, Zw 524-65). Three galaxies (Zw 525-21, Zw 540-65 and Zw 541-19)
certainly do not belong to the cluster because their velocity relative to
the cluster center is larger than 4000 km s$^{-1}$; they were eliminated
from the list of galaxies to be observed, together with all galaxies whose
morphological type is listed in Poulain, Nieto, Davoust (1992). Note
however that some of the observed galaxies might still be fore- or
background objects, in particular some of the galaxies for which the
radial velocity is unknown or some galaxies in the outskirts of the
studied region and with intermediate relative velocities with respect to
the cluster center. Figure 2 (only available in electronic form) shows the
spatial distribution of the galaxies in the studied region. 

54 of these galaxies were first inspected on Schmidt plate OCA \#2977
taken at the Calern Observatory in December 1992 and digitized, as the
preceding one, at the MAMA with a 5 micron (= 0.33{\tt "}) step and with
an aperture of 0.65{\tt "}. The remaining galaxies are outside the region
covered by our plate. All the galaxies were further inspected on the
Digitized Palomar Sky Survey\fonote{The Digitized Sky Survey was produced
at the Space Telescope Science Institute under US Government grant NAG
W-2166.}. Obvious spiral galaxies were classified as such and eliminated
from the list of galaxies to be observed in CCD. 

We found images of 3 galaxies in our CCD-image archives which had not been 
classified by Poulain, Nieto \& Davoust (1992); they were not reobserved.

Finally, we observed in CCD almost all galaxies not classified as S. Five
observing runs (35 nights) at TBL were used for completing this sample,
since bad weather, technical problems with the filter wheel and with data
aquisition, and the pear-shaped PSF of many images, made the completion of
the program very slow, largely compensating the good luck of Poulain,
Nieto \& Davoust (1992). During the first 3 runs (February 1993, December
1993, February 1994) we used a $1024\times1024$ Thompson CCD with a pixel
size of $19 \mu$ corresponding to 0.24 arcsec on the sky and the galaxies
were observed in the Gunn $r$ filter and calibrated in the Cousin $R$
filter, to make the measures consistent with those of Paper I. During the
last two runs (December 1994 and February 1995) the CCD was the same
Tektronics $1024\times1024$ as for the Coma sample and we observed the
galaxies in Johnson $V$. 

Thanks to the fact that the sample is largely composed of obvious spirals, 
whose large-scale spiral structure is visible even on defective images 
(missing pixels, pear-shaped PSF, unidentified filter), we were able to 
collect images of sufficient quality for the morphological classification of 
all but 4 galaxies (Zw 540-73, 540-80, 540-83 and 525-36). 

For the first three galaxies, we only used our Calern plates, whose densities 
had been transformed into intensities by means of a contour to contour 
correspondence between plate-image output and CCD-image intensity for a set of 
galaxies, as we did in Paper I.  Plate OCA \#2977, like those used for Coma 
galaxies, is a Kodak panchromatic 4415 emulsion sensitive from the near UV to 
6000--7000 \AA \ and was taken without filter. The fact that its sensitivity 
does not match the CCD $V$ or $r$ band prevented us from absolutely 
calibrating this plate, and therefore from computing passband dependent 
quantities (magnitudes, radius at a given surface brightness, etc.); but, for 
galaxies without pronounced color gradients, such as Es, we can still compute 
the density to intensity transformation (in arbitrary units) without 
significant errors. 

For Zw 525-36, we did not have any image of sufficient quality for
the morphological classification, and its type remains unknown.

Table 1 details the observing log. The median seeing was 1.4 arcsec, which 
corresponds to a restframe resolution of 0.66 Kpc, still comparable to the 
resolution available with the {\it Space Telescope} for distant galaxies. 

The method of analysis and the classification scheme are described in details 
in Paper I and references therein, and do not need to be 
presented anew.

\titlea {Results of the photometric and isophotal shape analyses}

\titleb {Spirals}

Table 2 presents the parameters of 31 galaxies in Coma classified as spirals 
from visual inspection of the two plates OCA \#2842 and OCA \#2849, with our 
notes and from the morphological appareance on CCD data when available. 
In that Table, we also list separately three Coma galaxies classified as S from 
CCD data taken on December 27 and 28, 1994, and two Coma galaxies classified 
as spirals from images in our archives.

\bigskip

Table 3 presents the parameters of 36 galaxies in Perseus classified as spiral 
or peculiar, from their visual appearance on plate OCA \#2977, on the 
Digitized Palomar Sky Survey and/or on our CCD images. 

\titleb {Early-type galaxies}

The data presented in Tables 4 and 5, for early-type galaxies in Coma and 
Perseus respectively, include the usual photometric parameters in 
Cousin's $R$ or Johnson $V$ band, namely the asymptotic magnitude, the 
effective radius, the corresponding isophotal major axis, and the average 
surface brightness (hereafter SuBr) inside the effective isophote.  
Geometrical parameters are given next, the minimum axis ratio (or 
alternatively its value at the effective isophote), the representative $e_4$ 
coefficient (see Paper I for the definition of $e_4$ and $f_4$), the axis 
ratio in the envelope, i.e. at the isophote $\mu_R$ = 24 mag arcsec$^{-2}$ or 
$\mu_V$ = 24.85 mag arcsec$^{-2}$, and a representative value for the 
isophotal twist. Next is a coded description, indicating the detection or 
absence of components such as bar, disk, spiral pattern, and the 
classification of disks and envelopes. 

We emphasize that the units of angular measures in this paper and in
Paper I are arcsec, not 0.1 arcmin as incorrectly stated in Paper I.

\bigskip

Table 4 lists the parameters of 28 early-type galaxies in Coma.
The first 4 columns concern the catalogue data, the others list the measured 
data.\par

Table 5 lists the parameters of 27 early-type galaxies in Perseus.
The first 2 columns concern the catalogue data, the others list the measured 
data.\par

In the notes to Table 4 and 5, we give qualitative remarks for galaxies
which present either peculiar morphological properties or practical
problems for classification.

\titlea {Quality of the parameters and types}

A detailed comparison of the morphological types of the whole Coma sample
of galaxies with other published studies is presented in Andreon \&
Davoust (1997). It shows that the main objective of this work and of Paper
I, {\it reliable estimates of morphological types}, has been reached,
since these types are at least as good as the traditional ones, because
less subjective, more reproducible and based on images of adequate
quality. 

The quality of the parameters listed in Tables 4 and 5 (magnitudes, effective 
radii, representative ellipticities, etc.) does not differ from that of 
the parameters presented in Paper I, because of the close similarity of the 
data and of the analyses. 

Some discrepancies have been found between our values of representative 
quantities (such as ellipticity or $e_4$) and published ones, but we stress 
that they are largely due to differences in the definition of what is a 
``representative" quantity, whether it is an intensity averaged quantity, or 
the quantity at the galaxy effective radius, at its maximum or at the 
extremum, and of what is its value when not just an extremum is present or 
when we only measure an incomplete range of galaxy radii (i.e. always because 
of seeing or sky brightness limitations).  In Paper I, the comparison of these 
``representative" quantities shows that the typical errors are of 0.06 on 
ellipticity (and our ellipticities are systematically larger than the others by 
0.05) and of 1.3 (\%) on $e_4$ (and our $e_4$ are larger than the others by 
0.7 \%). These figures, based on more than 200 comparisons, are also valid 
for the data presented in this paper. 

Aside from errors on sky determination, the effective radii suffer from
the existence of two definitions, the radius containing half the light,
measured by extrapolating the luminosity growth curve and taking the
radius where the integrated magnitude is 0.75 mag fainter than the total
one, or the slope of the SuBr profile, measured by the best fit of the
SuBr profile with a de Vaucouleurs' law.  Adopting the former method, the
subjective extrapolation of the growth curve implies a typical error of
0.02 in $\log(r_e)$ and $\log(l_e)$ for galaxies of range 0.5 to 1.0 in
$\log(r_e)$ (where $r_e$ is in units of arcsec), or, more precisely, this
is the typical scatter between estimates of different observers, all using
the same growth curves. Much larger differences have sometimes been found
for galaxies whose growth curves differ from the standard ones listed in
RC3 (de Vaucouleurs et al. 1991), used by us as standards. 
 
\titlea {Summary}

We present morphological type estimates, together with a detailed coded 
description for 59 galaxies in Coma and 80 in Perseus. The material for the 
morphological type estimates is adapted to the difficulty of morphological 
classification, ranging from Schmidt plates for obvious Ss to CCD data with 
a median restframe resolution of 0.65-0.75 Kpc for early-types. 

In the present paper and in two previous ones (Paper I; Poulain, Nieto \& 
Davoust 1992) we classify two magnitude complete samples of galaxies: all 
(187) galaxies in Coma brighter than $M_b=-19.2$ mag within 1 degree from the 
cluster center and all (139) galaxies in Perseus  brighter than $M_{Zwicky}=-
19.5$ mag in a box of $5\degr 3' \times 5\degr 27'$. At the distance of these 
two clusters, we sample an area of 17 and 91 Mpc$^2$ and distances of up to 
2.3 and 6.4 Mpc from the cluster center, for Coma and Perseus respectively, 
and 4 to 5 orders of magnitude in galaxy density. The ranges of explored 
clustercentric distances and galaxy densities allow us to study how the 
cluster affects the galaxy properties. The results for Coma, based on these 
data, are presented in Andreon (1996). For Perseus, work is in progress. 

\acknow{  
We thank Christian Pollas (Observatoire de Calern) for taking three
Schmidt plates for this project, and Jean Guibert and his staff (MAMA) for
digitizing the plates. We thank B\'en\'edicte Rougeaux for her visual
inspection of the OCA \#2849 plate. This research made use of the
NASA/IPAC Extragalactic Database (NED) which is operated by the Jet
Propulsion Laboratory, California Institute of Technology, under contract
with the National Aeronautics and Space Administration and of the
Lyon-Meudon Extragalactic Database (LEDA) supplied by the LEDA team at the
CRAL-Observatoire de Lyon (France). Skyeye
(http://terra.ira.bo.cnr.it/ira/skyeye) greatly helped giving a quick and
easy access to the Digitized Palomar Sky Survey}

\begref {References}

\ref Andreon, S. 1994, A\&A 284, 801

\ref Andreon S., 1996, A\&A, 314, 763

\ref Andreon S., Davoust E., 1997, A\&A, in press

\ref Andreon S., Davoust E., Heim T., 1997, A\&A, in press

\ref Andreon S., Davoust E., Nieto J.-L., Michard R., Poulain P., 1996,
A\&AS, 116, 429 (Paper I)

\ref Bucknell M., Godwin J., Peach J., 1979, MNRAS 188, 579

\ref Butcher H., Oemler A., 1985, ApJS 57, 665

\ref Chincarini G., Rood H., 1971, ApJ 168, 321 

\ref Dressler A., 1980, ApJS 42, 565

\ref Dreyer J., 1888, {\it The New General Catalogue of nebulae and of
clusters of stars} et {\it Index Catalogue}, Memoirs
of the Royal Astronomical Society vol. 49,51,59

\ref Gavazzi G., Boselli A., Carrasco L. et al., 1995, A\&AS 112, 257 (GavIV)
 
\ref Gavazzi G., Garilli B., Boselli A., 1990, A\&AS 83, 399 (GavI)

\ref Gavazzi G., Randone I., 1994,  A\&AS 107, 285 (GavIII)

\ref Godwin, J.G., Metcalfe, N., Peach, J.V. 1977, MNRAS 202, 113 (GMP)

%\ref Hubble E., 1936, {\it The Real of the Nebulae},  New Haven: Yale 
%University Press

\ref Kent S., Sargent W., 1983, AJ 88, 697 

\ref Ledlow M, Owen M., 1995, AJ 110, 1959

\ref Michard 1996,  Astrophys. Lett. \& Comm., 31, 187

\ref Michard, R., Marchal, J. 1993, A\&AS, 98, 29
 
\ref Michard, R., Marchal, J. 1994a, A\&AS, 105, 481 
 
\ref Michard, R., Marchal, J. 1994b, A\&A, 283, 779

\ref Nilson P., 1973, {\it Uppsala General Catalogue}, 
Uppsala Astr. Obs. Annals, vol.6  (UGC)

\ref Poulain P., Nieto J.-L., Davoust E., 1992, A\&AS 95, 129

\ref Rood, H., Baum, W., 1967, AJ 72, 398

\ref Shombert J., 1987, ApJS 64, 643

\ref Tiff W., 1977, ApJ 222, 54

\ref Tonry J., 1987, in {\it Structure and Dynamic of
Elliptical Galaxies}, ed. T. de Zeeuw (Dordrecht: Reidel), IAU
Symp. 127, p. 89

\ref Zwicky F. et al., 1961-1968, {\it Catalogue of Galaxies and Clusters
of Galaxies}, (Pasadena: Caltech)

\endref

\vfill\eject

%\vfill\eject\null\vfill\eject

\psfig{figure=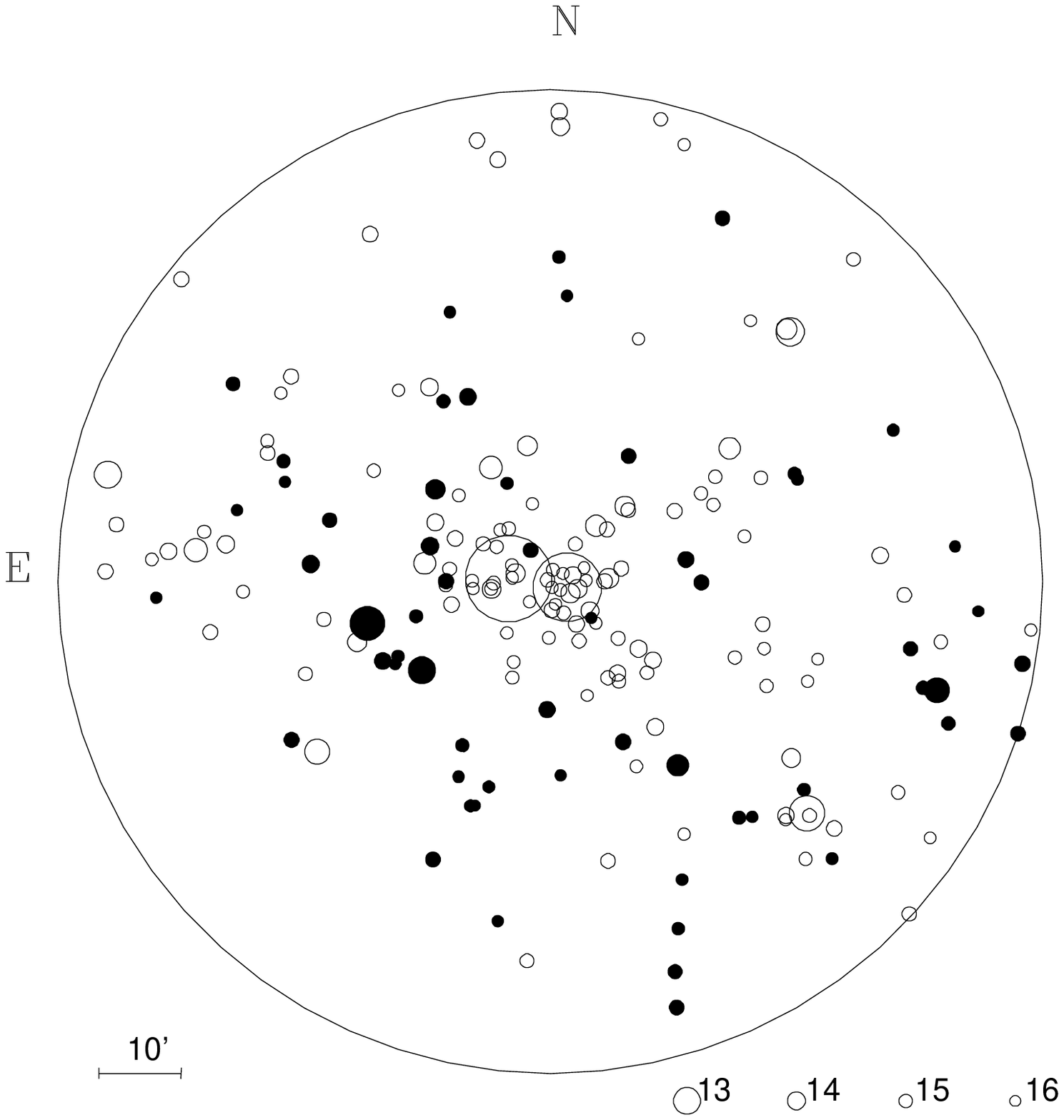,height=18truecm}

\noindent {\bf Figure 1}. The galaxies of the Coma sample. Full and open
circles represent galaxies whose morphological type is presented in this
paper and in Paper I, respectively. The size of the circles is
proportional to the magnitude of the galaxy. The radius of the field is
one degree.  There are 187 galaxies in this field. The morphological types
of 59 of them are presented in this paper, and the others can be found in
Paper I. 
\vfill\eject
\null\vfill\eject
%\vfill\eject

\psfig{figure=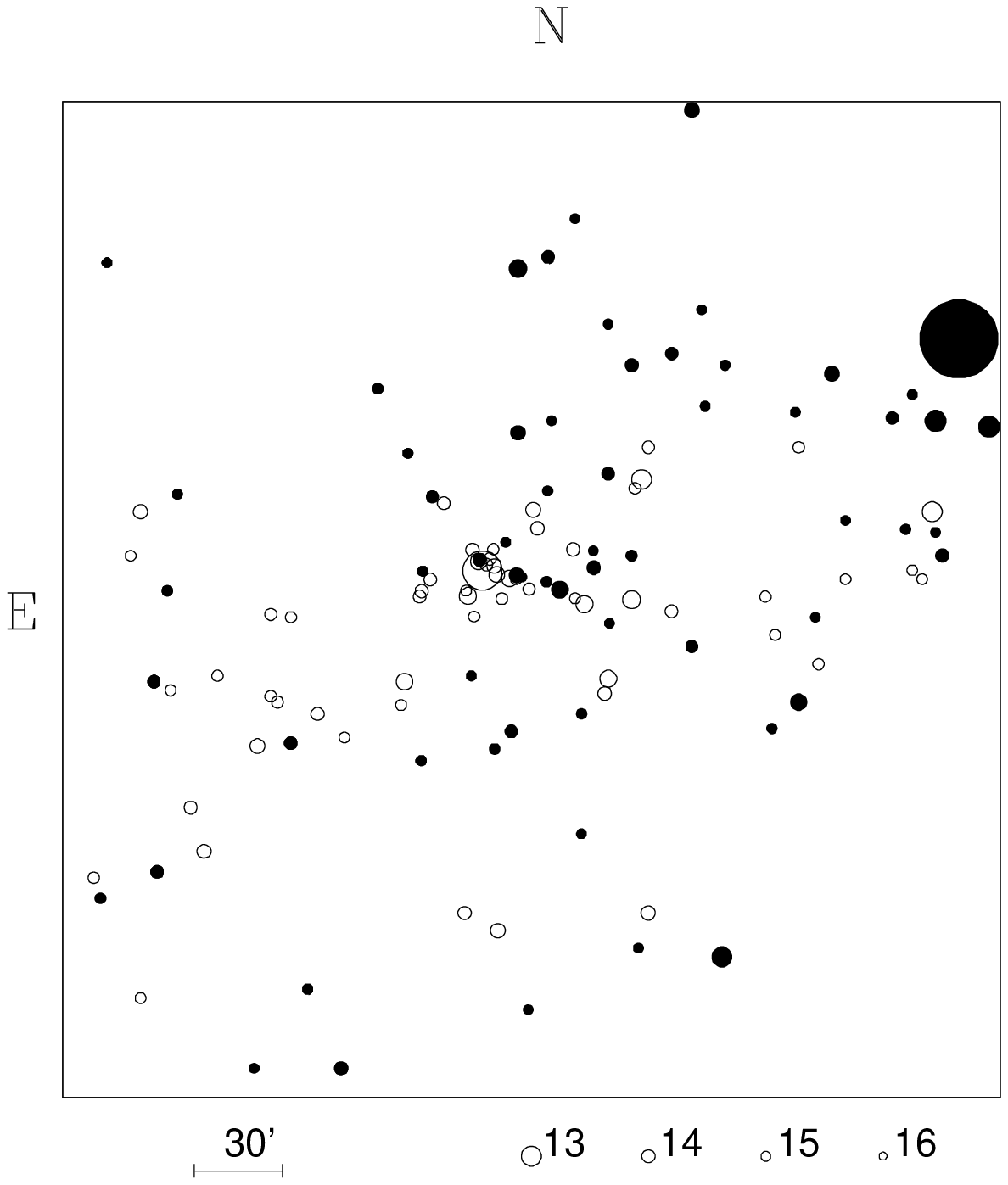,height=18truecm}

\noindent
{\bf Figure 2}.  The galaxies of the Perseus sample. Full and open circles
represent galaxies whose morphological types are presented in this paper
and in Poulain, Nieto \& Davoust (1992), respectively. The size of the
circles is proportional to the magnitude of the galaxy. The size of the
studied region is $5\degr 3' \times 5\degr 27'$. There are 139 galaxies in
this field. The morphological types of 80 of them are presented in this
paper, and the others can be found in Poulain, Nieto \& Davoust (1992).

\vfill\eject
\null\vfill\eject
%\vfill\eject

\baselineskip=9.95 truept

\tabcap{1}{Observing log of Perseus galaxies}
\halign{#\hfill&\hfill#\hfill&\hfill#\hfill&\quad\hfill#&\quad\hfill#&\quad#\hfill\cr
\noalign{\medskip\hrule\medskip}
name  & run& filter& exp. time& seeing& notes\cr 
        & & & (min)& (arcsec)&\cr
\noalign{\medskip}
525-19   & dec94& V& 15& 1.7& \cr
525-29   & dec94& x& 15& 1.6& \cr
525-31   & feb95& V& 15& 1.5& \cr
525-6    & dec94& x& 15& 1.6& \cr
540-32   & feb95& V& 15& 1.3& \cr
540-33   & feb95& V& 15& 1.3& \cr
540-35   & dec94& V& 15& 1.6& \cr
540-36   & dec94& V& 15& 1.7& \cr
540-41   & feb95& V& 15& 1.6& \cr
540-42   & dec94& x& 15& 1.6& \cr
540-43   & dec94& x& 15& 1.6& \cr
540-45   & dec94& V& 15& 1.7& \cr
540-46   & sep89& r& 15& 1.3& archive, PIC, CCD=RCA1 F/10 pixel size=0.324 \cr
540-48   & feb95& V& 15& 1.2& \cr
540-51   & feb95& V& 15& 1.4& \cr
540-54   & dec94& x& 15& 1.5& \cr
540-56   & dec93& r& 20& 2.6& \cr
540-57   & feb95& V& 15& 1.3& \cr
540-59   & dec93& r& 20& 3.7& \cr
540-60   & dec94& x& 15& 1.6& \cr
540-62   & feb95& V& 15& 1.2& \cr
540-67   & dec93& r& 20& 2.6& \cr
540-68   & dec94& V& 15& 1.7& \cr
540-70   & dec94& x& 15& 1.6& \cr
540-73   & plate&  &   &    & OCA \#2977 \cr
540-74   & feb94& r& 20& 1.0& \cr
540-76   & feb94& r& 20& 1.2& \cr
540-80   & plate&  &   &    & OCA \#2977 \cr
540-81   & dec89& r& 10& 0.7& archive, CFH, CCD=RCA2 (see Paper I)\cr
540-82   & dec94& V& 15& 1.6& \cr
540-83   & plate&  &    &    & OCA \#2977 \cr
540-85   & feb95& V& 15& 1.3& \cr
540-85   & feb95& V& 15& 1.6& \cr
540-89   & feb95& V& 15& 1.4& \cr
540-97   & dec93& r& 20& 2.5& \cr
540-104  & plate&  &   &    & OCA \#2977 \cr
540-106  & feb95& V& 15& 1.5& \cr
540-114  & dec93& r& 20& 2.7& \cr
540-114  & plate&  &   &    & OCA \#2977 \cr
540-115  & dec94& V& 15& 1.5& \cr
541-3    & dec93& r&  2& 1.7& \cr
541-12   & dec93& r& 20& 1.8& \cr
541-14   & dec93& r& 20& 1.9& \cr
541-15S  & sep90& r& 30& 1.1& archive, PIC, CCD=Tho F/10 pixel size=0.2?\cr
541-16N  & feb95& V& 15& 1.4& \cr
541-16S  & feb95& V& 15& 1.4& \cr
541-112N & dec93& r& 20& 2.7& \cr
BGP41    & feb94& r& 15& 1.1& \cr
 T40     & feb95& V& 15& 1.4& \cr
 T40     & feb95& V& 15& 1.3& \cr
 T04     & dec93& r& 20& 1.8&  Not analyzed, bright star at 14"\cr
\noalign{\medskip\hrule\medskip}
}
\noindent
Notes : \hfill\break 
feb93: CCD Thompson 1024x1024, pixel size=0.24 \hfill\break
dec93: CCD Thompson 1024x1024, pixel size=0.48 (0.24 2x2 binned) \hfill\break
feb94: CCD Thompson 1024x1024, pixel size=0.24 (the observations
in the x filter were done with an unknown filter, because of
a mechanical fault). \hfill\break
dec94: CCD Tektronics 1024x1024, pixel size=0.30 \hfill\break
feb95: CCD Tektronics 1024x1024, pixel size=0.30 \hfill\break

\vfill\eject
\null\vfill\eject

\tabcap{2}{Galaxies classified as S from visual inspection of the plates 
OCA2842 and OCA2849 (all but two not presented in Paper I)}
\halign{#&\quad#\hfill&\quad#\hfill&\quad#\hfill\cr
\noalign{\medskip\hrule\medskip}
GMP$_a$ &       characteristics                 &other observations     &notes\cr
\noalign{\medskip}
0315&   irr                     &run 10, S              &     \cr  
0433&   asym                    &                       &dust, edge-on, uncertain type \cr
0440&   irr, asym               &                       &                            \cr
0507&   irr, asym               &run 9, S               &Irr\cr
0510&   irr, asym, low SuBr     &                       &Irr\cr
0686&   arms                    &run 7, S, GavIII, S    &\cr
0689&   asym, irr               &                       &\cr
0790&   irr, arms               &run 10, S              &\cr
0804&   irr, arms               &GavI, S, GavIV, S      &beautiful S \cr
0834&   irr?, spiP?             &                       &uncertain type, classified S by D80, BO \cr
0837&   arms                    &GavI, S                &beautiful barred S                   \cr
0867&   irr                     &run 9, SA0/a           &strange shape on plate
\cr
0875&   irr, asym               &                       &late type spiral or Irr                       \cr
0892&   asym                    &                       &dust, no bulge \cr
0897&   irr                     &                       &Irr          \cr
0914&   asym, arms              &run 10, S              &\cr
0978&   asym, irr               &                       &\cr
1001&   irr                     &run 10, S              &outside the redshift limits\cr
1193&   irr, asym               &                       &\cr
1203&   asym                    &                       &\cr
1420&   irr, asym               &GavIV, S               &\cr
1555&   irr                     &GavIV, S               &Irr\cr
1566&   irr, asym               &                       &Irr\cr
1576&   arms                    &                       &\cr
1675&   irr, asym               &GavIII, S              &\cr
1711&   spiP                    &                       &beautiful barred S\cr
1744&   spiP                    &                       &uncertain type, classified S by BO\cr
2156&   irr                     &                       &Irr                            \cr
2172&   irr, asym, spiP?        &                       &late type spiral\cr
2275&   asym                    &                       &dust              \cr
\noalign{\medskip\hrule\medskip}
0315&   asym \cr
0790&   spiP  \cr
1001&   irr, asym\cr
\noalign{\medskip\hrule\medskip}
0507&   arms, irr&              march 94&               Irr, not in the published list\cr
0518&   S in kp1608&    run 7, S&  not published in Paper I\cr
\noalign{\medskip\hrule\medskip}
}
\noindent
Notes:\hfill\break
BO= Butcher \& Oemler 1985\hfill\break
GavI= Gavazzi, Garilli \& Boselli 1990\hfill\break
GavIII= Gavazzi \& Randone 1994\hfill\break
GavIV= Gavazzi, Boselli \& Carrasco 1994\hfill\break
D80= Dressler 1980\hfill\break
kp1608= KPNO photographic plate (see Paper I)\hfill\break
run 7= CCD observations in March 1993 (see Paper I)\hfill\break
\vfill\eject
\null\vfill\eject

\tabcap{3} {Morphological description of spiral galaxies in Perseus}
\halign{#\hfill\ &#\hfill\ &#\hfill\ &#\hfill\ &#\hfill\ &#\hfill\ &#\hfill\ &#\hfill\ \cr
\noalign{\medskip\hrule\medskip}
 Name       & Palomar dig.        & OCA \#2977  & Notes OCA \#2977                  & CCD                         & Run n. \& Notes                  \cr
\noalign{\medskip}
 Zw 525-6    &                     &          &                                & S                           & dec94                  \cr
 Zw 525-11   &                     & S        & very low SuBr                  &                             &                        \cr
 Zw 525-19   &                     &          &                                & (r)S barred                     & dec94                  \cr
 Zw 525-29   &                     &          &                                & S                           & dec94                  \cr
 Zw 525-31   &                     &          & too close to bright star & S                           & feb95                  \cr
 Zw 525-33   & Irr & & & & \cr
 Zw 525-39   & S & & & & \cr
 Zw 540-34   & S arms              &          &                                &                             &                        \cr
 Zw 540-35   &                     &          &                                & S                           & dec94                  \cr
 Zw 540-36   &                     &          &                                & S flocculent                & dec94                  \cr
 Zw 540-37   & S late              &          &                                &                             &                        \cr
 Zw 540-42   &                     &          &                                & S                           & dec94                  \cr
 Zw 540-43   &                     &          &                                & S asymm                     & dec94                  \cr
 Zw 540-45   &                     &          &                                & S                           & dec94                  \cr
 Zw 540-49   & S dust              & S        &                                &                             &                        \cr
 Zw 540-54   &                     &          &                                & S barred                        & dec94                  \cr
 Zw 540-58   & S irr               & Pec      & Interact                       &                             &                        \cr
 Zw 540-60   &                     &          &                                & 
S late                      & dec94                  \cr
 Zw 540-67   &                     &          &                                & Irr                         & dec93                  \cr
 Zw 540-68   &                     &          &                                & S                           & dec94                  \cr
 Zw 540-70   &                     & late type?&                                & S dusty                     & dec94                  \cr
 Zw 540-71   &                     & S        & low SuBr, asym                 &                             &                        \cr
 Zw 540-77   &                     & S        &                                &                             &                        \cr
 Zw 540-82   &                     &          &                                & S                           & dec94                  \cr
 Zw 540-84   &                     & Irr      &                                &                             &                        \cr
 Zw 540-90   & S late              & S        &                                &                             &                        \cr
 Zw 540-91   & S arms              & S        &                                &                             &                        \cr
 Zw 540-93   & S                   &          &                                &                             &                        \cr
 Zw 540-94   &                     & S        & low SuBr, elongated, -$>$dust   &                             &                        \cr
 Zw 540-97   &                     &          &                                & S two faint arms, bar, disk & dec93                  \cr
 Zw 540-106  & S                   & disk     & disk                           &                             & feb95                   \cr
 Zw 540-115  &                     &          & other galaxy at 6"             & S dusty                     & dec94                  \cr
 Zw 540-118  & S?                  & S        &                                &                             &                        \cr
 Zw 540-121  &                     & S        & disk, irr                      &                             &                        \cr
 Zw 541-3    & Irr    &          &                                &          Irr                   &  dec93, embedded ]\cr
& & & & & in the halo of ] \cr
& & & & & a bright star, \cr
 Zw 541-19   & S arms, HII reg.    &          &                                &                             & out of redshift ]\cr
& & & & & limits \cr
 Zw 541-112N &                     &          &                                & Pec (Interact?)              & dec93                  \cr\noalign{\medskip\hrule\medskip}
}

\vfill\eject
\null\vfill\eject

{\bf Description of Table 4 (Coma galaxies)}
\bigskip
  
\noindent (1) and (2) Number in the abridged and unabridged versions of the 
Godwin, Metcalf \& Peach (GMP) catalogue. 

\noindent (3) Number in Dressler's catalogue (Dressler 1980).

\noindent (4) Usual designation, such as NGC, IC (Dreyer 1888), and RB (Rood 
\& Baum 1967) numbers. 

\noindent (5) Asymptotic magnitude, in V (Johnson's system).
 
\noindent (6) Logarithm of the effective radius, in units of arcsec ($\log(r_e)$).
 
\noindent (7) Logarithm of the semi major axis of the effective isophote, in 
units of arcsec ($\log(l_e)$). 
 
\noindent (8) Mean SuBr inside the effective isophote, in V mag arcmin$^{-2}$. 
 
\noindent (9) Photometric evidence for a disk, coded as st (strong), cl (clear),
ft (faint), or no (none).
 
\noindent (10) Typical axis ratio, either its minimum value, if clearly defined,
or its value at the effective isophote otherwise.
 
\noindent (11) Location where the axis ratio was estimated, coded as ex (at its 
extremum), re (at the effective isophote), or co (if the value is the same at 
both locations).

\noindent (12) Typical $e_4$ parameter, either its extremum value, if clearly 
defined, or its value at the effective isophote otherwise. The estimates are 
in \%.  
 
\noindent (13) Location where the $e_4$ parameter was estimated, coded as ex 
(at its extremum), re (at the effective isophote), or co (if the value is the 
same at both locations). 

\noindent (14) Axis ratio in the envelope, i.e. at the isophote $\mu_V$ = 
24.85 mag arcsec$^{-2}$. 
 
\noindent (15) Amplitude of isophotal twist in the range of reliable 
measurements, in degrees. 

\noindent (16) Detection of a bar, coded as follows : bar (bar seen), bar? (bar 
suspected), -no (no bar seen). 
 
\noindent (17) Detection and classification of a disk, coded as follows: emDi 
(embedded disk), miDi (mixed disk), exDi (extended disk), -?Di (detected but 
unclassified disk), -no- (no disk seen). 
 
\noindent (18) Detection of a spiral pattern, coded as follows: spiP  (spiral 
pattern seen), spiP? (spiral pattern suspected), -no-  (no spiral pattern 
seen). 
 
\noindent (19) Classification of an envelope, coded as follows: spH 
(spheroidal halo), thD (thick disk), exD (extended disk), pec (peculiar 
envelope), -?- (unclassified envelope). 
 
\noindent (20) Our morphological classification, coded as follows: boE (boxy 
E), unE (undetermined E), diE (disky E), SA0, SAB0, SB0, Sa, etc., S... 
(spiral of unknown stage). 
 
\noindent (21) An asterisk refers to a note in Sect. 5 about specific features 
such as important dust pattern, ring or lens, low SuBr, $f_4$-asymmetry, etc., 
and about uncertainties of various sources. 
\smallskip 

Note that, when a parameter has not been measured, or when a specific 
morphological component has not been studied, the relevant code is replaced by 
a dash. 

\vfill\eject\null\vfill\eject

\tabcap{4}{Morphological description of early-type galaxies in Coma}
\halign{#\hfill\ &#\hfill\ \ &\hfill\ \ #&\hfill\ #&\quad\hfill\ #&\hfill\ #&\hfill\ #&\hfill\ #&\hfill\ #&\hfill\ #&\hfill\ #&\hfill\ #&
\hfill\ #\ \hfill& \hfill\ #&\hfill\ #\ \hfill&\hfill\ #&\hfill\ #&\hfill\ #&\hfill\ #\ &#\hfill\ &#\hfill\ \cr
\noalign{\medskip\hrule\medskip}
(1)& (2)& (3)& (4)& (5)& (6)& (7)& (8)& (9)& (10)& (11)& (12)& (13)& (14)& (15)& (16)& (17)& (18)& (19)& (20)& (21)\cr
0552 & 1715&  141& NGC 4927& 13.56& 0.93& 1.02& 11.31& no& 0.70& co& 1.2& co& 0.73&  3& -no & miDi& -no-&  thD& diE  &* \cr    
0607 & 1853&  190&      & --.--& -.--& -.--& --.--& st& 0.26& ex& 8.0& ex& 0.29&  2& -no & miDi& -no-&  thD& SA0  &* \cr 
0718 & 2157&   79& NGC 4919& 13.73& 0.71& 0.80& 10.39& st& 0.72& re& 2.2& ex& 0.49&  4& -no & -?Di& -no-&  pec& SA0  &* \cr
0750R& 2237&   80&      & 14.63& 0.55& 0.65& 10.47& st& 0.59& ex& 3.0& co& 0.99&  2& -no & emDi& -no-&  spH& SA0  &* \cr
0750 & 2237&   80&      & 15.06& 0.55& 0.65& 10.91& st& 0.59& ex&  3.0& co& 0.92& 30& -no & emDi& -no-&  spH& SA0  &* \cr
0754 & 2259&  229&      & 15.01& 0.59& 0.64& 11.03& no& 0.80& re& 1.8& co& 0.71& 10& -no & miDi& -no-&  exD& diE  &* \cr
0828 & 2417&  167& NGC 4908& 13.77& 0.70& 0.78& 10.39& ft& 0.67& ex& -0.8& ex& 0.65&  8& -no & -no-& -no-&  -?-& unE  &* \cr
0857 & 2495&  231&      & 14.55& 0.42& 0.54&  9.73& cl& 0.50& ex& 4.0& ex& 0.60&  0& -no & miDi& -no-&  spH& SA0  &  \cr
0908 & 2629&  232& NGC 4896& 13.77& 0.87& 1.00& 11.20& ft& 0.52& ex& 0.0& re& 0.55&  3& -no & -no-& -no-&  -?-& unE  &* \cr
0924 & 2670&   27&      & 15.26& 0.51& 0.57& 10.90& ft& 0.80&  ex& 1.6& ex& 0.82&  3& -no & -?Di& -no-&  -?-& diE  &  \cr
0967 & 2776&   39&      & 14.91& 0.54& 0.58& 10.73& no& 0.78&  ex& 0.0& co& 0.78&  2& -no & -no-& -no-&  -?-& unE  &* \cr
1109 & 3073&  175& NGC 4883& 14.24& 0.61& 0.64& 10.44& cl& 0.71& ex& -1.0& re& 0.82& 45& bar & -?Di& -no-&  -?-& SB0  &* \cr
1154 & 3165&   57&      & 13.85& 0.65& 0.81& 10.22& st& 0.25&  ex& 8.0& ex& 0.34&  4& -no & emDi& -no-&  spH& SA0  &* \cr
1232 & 3328&  242&      & 15.01& 0.36& 0.44&  9.91& cl& 0.50&  ex& 2.5& ex& 0.52&  2& -no & miDi& -no-&  thD& SA0  &  \cr
1432 & 3818&  218&      & 14.18& 0.60& 0.70& 10.26& no& 0.45&  ex& 2.5& co& 0.45& 11& -no & -?Di& spiP&  exD& S..  &* \cr
1560 & 4147&     &      & 14.27& 0.64& 0.74& 10.59& st& 0.50&  ex& 3.2& co& 0.61&  1& -no & emDi& -no-&  spH& SA0  &  \cr
1564 & 4156&   43& NGC 4853& 13.49& 0.54& 0.58&  9.29& st& 0.78&  ex& 0.0& re& 0.82& 10& -no & -no-& -no-&  -?-& SAB0p&* \cr
1594 & 4230&  161& RB 241& 13.98& 0.76& 0.80& 10.89& no& 0.87& co& -0.8& co& 0.91& 20& -no & -no-& -no-&  -?-& boE  &* \cr
1625 & 4315&  137& NGC 4850& 14.23& 0.58& 0.65& 10.24& cl& 0.74& co& 3.1& co& 0.93& 30& bar & -?Di& -no-&  -?-& SB0  &* \cr
1844 & 4849&  211&      & 14.31& 0.78& 1.01& 11.30& st& 0.31&  ex& 7.1& co& 0.44&  5& -no & emDi& -no-&  spH& SA0  &* \cr
1852 & 4866&  212&      & 14.88& 0.67& 0.77& 11.32& cl& 0.50&  ex& 1.4& co& 0.56& 12& bar?& -?Di& -no-&  thD& SAB0 &* \cr
1925 & 5038&   16&      & 15.11& 0.67& 0.73& 11.54& cl& 0.72&  co& 4.1& co& 0.89& 60& -no & emDi& -no-&  pec& SA0  &* \cr
2047 & 5341&  221&      & 14.98& 0.47& 0.55& 10.44& cl& 0.64&  ex& 4.9& co& 0.95& 74& bar & -?Di& -no-&  -?-& SB0  &  \cr
2085 & 5428&     &      & 14.46& 0.64& 0.66& 10.74& no& 0.91&  co& -0.8& co& 0.90&  9& -no & -no-& -no-&  -?-& boE  &* \cr
2109 & 5495&     &      & 14.51& 0.47& 0.51&  9.94& no& 0.79&  ex& 1.0& ex& 0.88& 15& -no & -?Di& -no-&  -?-& unE  &* \cr
2134 & 5568&     & NGC 4816& 12.66& 1.40& 1.44& 12.74& no& 0.79& re& -1.0& ex& 0.74& 20& -no & -no-& -no-&  -?-& boEp &* \cr 
2220 & 5799&     &      & 15.14& 0.38& 0.44& 10.13& cl& 0.69&  ex& 2.9& co& 0.81& 30& bar & -?Di& -no-&  -?-& SB0  &  \cr
2283 & 5999&     &      & 14.03& 1.00& 1.09& 12.13& no& 0.78&  ex& 0.0& co& 0.80&  0& -no & -no-& -no-&  -?-& unE  &* \cr
\noalign{\medskip\hrule\medskip}
}

\vfill\eject\null\vfill\eject

{\bf Notes to Table 4 (Coma galaxies)}
\bigskip

\noindent {\bf 0552}: dust along the major axis. Type given in Paper I 
confirmed. 
  
\noindent {\bf 0718}: two bumps on the major axis profile. Inner ring and lens.
Type given in Paper I confirmed.
 
\noindent {\bf 0750R}: image of run 9 (see Paper I).
 
\noindent {\bf 0750}: 30 degree twist at r $\sim 10$ arcsec on the V image. 
same twist in R, but image not deep enough for full confirmation. Type given in 
Paper I confirmed. 
 
\noindent {\bf 0754}: type given in Paper I fully confirmed.
 
\noindent {\bf 0828}: E boxy outside $r_e$ and possibly disky inside. In Paper 
I it is classified diE because the boxiness of the outer region is not 
detected with certainty, partly because of the presence of a companion galaxy 
at 15 arcsec in the minor axis direction. 
 
\noindent {\bf 0857}: the bar suspected in Paper I is not detected in our 
better images. 
 
\noindent {\bf 0908}: slightly boxy isophotes out to $r_e$, then slightly 
disky isophotes. Uncertain detailed classification. 
 
\noindent {\bf 0967}: focus problems made the PSF slightly elongated in the NS 
direction and galaxy data inside 3 arcsec not usable. 
 
\noindent {\bf 1109}: the bar suspected in Paper I is detected.
 
\noindent {\bf 1154}: asymmetric with respect to its major axis.
 
\noindent {\bf 1432}: classified SA0/a by morphologists, probably because of 
the low contrast spiP. 
 
\noindent {\bf 1564}: ellipticity and $e_4$ profile unusual for a lenticular 
galaxy. It seems to be asymmetric. 
 
\noindent {\bf 1594}: the better data with respect to Paper I allow us to 
classify it elliptical. 
 
\noindent {\bf 1625}: the bar suspected in Paper I is detected.
 
\noindent {\bf 1844}: isophote twist toward GMP 1852. Type given in Paper I 
confirmed. 
 
\noindent {\bf 1852}: isophote twist opposite to GMP 1844, bar suspected but 
nearby star slightly elongated in the same direction. Type given in Paper I 
confirmed. 
 
\noindent {\bf 1925}: large isophote twist at low SuBr.
 
\noindent {\bf 2085}: rich in globular clusters\fonote{The expression 
``globular cluster" here and hereafter does not imply a group of dynamically 
bound stars, but a clump of light.}. The better data with respect to Paper I  
allow us to classify it elliptical. 
 
\noindent {\bf 2109}: focus problems made the PSF slightly elongated in a 
direction orthogonal to the galaxy major axis. Data inside 4 arcsec are not 
usable. Also classified unE with CFH data of Paper I. 
 
\noindent {\bf 2134}: rich in globular clusters. Its intensity profile obeys 
the $r^{-1.44}$ law from 1 arcsec out to 60 arcsec, putting this galaxy in the 
D class following Schombert (1987) and Tonry (1987). Classified as unE in 
Paper I from a slightly shallower image. 

\noindent {\bf 2283}: the analysis is made difficult by the crowded field and a 
nearby saturated star.

\vfill\eject
\null\vfill\eject

{\bf Description of Table 5 (Perseus galaxies)}
\bigskip

\noindent (1) Zwicky name.
 
\noindent (2) Usual designation, such as NGC, UGC (Nilson 1976), T (Tiff 1976), 
BGP (Bucknell, Godwin \& Peach 1979)
and CR (Chincarini and Rood 1977) numbers. 
 
\noindent (3) Flag for filter or observational material. 
 
\noindent (4) Asymptotic magnitude.
 
\noindent (5) Logarithm of the effective radius, in units of arcsec.
 
\noindent (6) Logarithm of the semi major axis of the effective isophote, in 
units of arcsec. 

\noindent (7)  Mean SuBr inside the effective isophote, in mag arcmin$^{-2}$.

\noindent (8) to (12) Same as columns (9) to (13) of Table 4.

\noindent (13) Axis ratio in the envelope, i.e. at the isophote V=24.85 or R=24 
mag arcsec$^{-2}$.
 
\noindent (14) to (20) Same as columns (15) to (21) of Table 4.
\smallskip 

Note that a ``(1)" in the table means that the parameter has not been computed 
because the night was not photometric, or because of the pan-chromaticity of 
the plate. 

\tabcap{5}{Morphological description of early-type galaxies in Perseus}
\halign{#\hfill\ &#\hfill\ \ &\hfill\ \ #&\quad\hfill\ #&\hfill\ #&\hfill\ #&\hfill\ #&\hfill\ #&\hfill\ #&\hfill\ #&\hfill\ #&\hfill\ #&
\hfill\ #&\hfill\ #&\hfill\ #&\hfill\ #&\hfill\ #&\hfill\ #\ &#\hfill\ &#\hfill\ \cr
\noalign{\medskip\hrule\medskip}
(1)& (2)& (3)& (4)& (5)& (6)& (7)& (8)& (9)& (10)& (11)& (12)& (13)& (14)& (15)& (16)& (17)& (18)& (19)& (20)\cr
\noalign{\medskip}
540-32& NGC 1175&  V        &  13.81& 0.90& 0.92& 11.41& no& 0.92& ex& -0.8& ex& 0.92:& 15:& -no& -no-& -no-& -no& boE& *\cr
540-33& NGC 1177&  V        &  12.52& 1.04& 1.19& 10.79& st& 0.27& ex& -7.2& co& 0.30:&   0& -no& emDi& -no-& thD& SA0& *\cr
540-41&      & V         &  14.28& 0.86& 0.94& 11.66& st& 0.62& co&  5.3& co&   1.0&  20& bar& emDi& -no-& spH& SB0& *\cr
540-46& UGC 2559& R        &    (1)& 0.95& 1.03&   (1)& st& 0.39& ex&  5.0& ex&   (1)&  10& bar& -?Di& -no-&  -?-& SB0& *\cr
540-48&      & V         &  14.41& 0.56& 0.66& 10.31& st& 0.52& ex&  3.0& ex&  0.63&   9& -no& miDi& -no-& thD& SA0& \cr
540-51&      & V        &  14.10& 0.77& 0.83& 11.06& st& 0.58& ex&  0.0& re&  0.65&  25& bar& -?Di& -no-& thD& SB0& \cr
540-56&      & R        &    (1)&   --&   --&   (1)& st& 0.39& ex&  7.0& ex&   (1)&   3& -no& miDi& -no-& thD& SA0& \cr
540-57&      & V         &  13.55& 0.78& 0.84& 10.54& no& 0.62& ex& -2.0& ex&  0.70&  10& -no& -no-& -no-&  -?-& boE& *\cr
540-59&      & R        &    (1)& 0.72& 0.78&   (1)& st& 0.46& ex&  5.6& ex&   (1)&  12& bar& -?Di& -no-& thD& SB0& \cr
540-62&      & V         &  13.40& 0.74& 0.82& 10.22& no& 0.67& ex& -1.0& co& 0.76:&   0& -no& -no-& -no-&  spH& boE& \cr
540-73&  CR 6 & plate    &    (1)&  -- & --  &   (1)&   &     &  &      & &    &&   &     &     &    & SA0& *\cr
540-74& BGP 37& R        &    (1)& 0.75& 0.80&   (1)& st& 0.75& co&    0& co&   (1)&   3& -no&   --& -no-& thD& SA0& *\cr
540-74& BGP 37& plate    &    (1)& 0.76& 0.83&   (1)& ft& 0.85& co&    0& co&   (1)&  10& -no&   --& -no-& -?-& SA0& *\cr
540-76&      & R        &       &     &     &      &   &     &    &    &   &      &    &    &     &     &    & SA0& *\cr
540-80&      & plate    &  14.38& 0.81& 0.89& 11.55& st& 0.75& ex&    0& co&  0.75&  25& bar& -?Di& -no-&  -?-& SB0& \cr
540-81&      & R        &       &     &     &       &   &    &  &       & &    && &       &   &      & SA0& *\cr
540-83& BGP 34& plate    &    (1)&  -- & --  &   (1)& cl& 0.23& ex&  9.0& ex&   (1)&   0& -no& emDi& -no-& thD& SA0& \cr
540-85& CR 15 & 2 V images &  13.91& 0.68& 0.73& 10.40& no& 0.72& co&  0.3& co&  0.76&   3& -no& -no-& -no-&  -?-& diE& \cr
540-89& CR 22 & V        &  14.06& 0.88& 1.03& 11.54& cl& 0.42& ex&  3.0& ex&    --&  15& -no&   --& -no-&  -?-& SA0& *\cr
540-106&     & V         &  14.24& 0.58& 0.73& 10.25& st& 0.32& ex&  6.1& ex&    --&  10& -no& emDi& spiP&  -?-& S  & *\cr
540-114&     & R        &    (1)&   --&   --&   (1)& st& 0.31& ex&  8.3& ex&   (1)&   0& -no& miDi& -no-& thD& SA0& *\cr
540-114&     &plate     &    (1)& 0.72& 0.88&   (1)& st& 0.35& ex&  6.9& ex&   (1)&   0& -no& miDi& -no-& exD& SA0& *\cr
541-12&      & R        &    (1)& 0.71& 0.86&   (1)& st& 0.31& ex&  8.1& ex&   (1)&   2& -no& emDi& -no-& spH& SA0& \cr
541-14&      & R        &    (1)& 0.93& 1.04&   (1)& st& 0.56& ex&  5.2& ex&   (1)&  25& bar& -?Di& -no-& spH& SB0& \cr
541-15S& UGC 2755&  V       &    (1)& 0.97& 0.99&   (1)& ft& 0.75& ex&  2.0& ex&   (1)&  60& bar& -?Di& -no-&  -?-& SB0& *\cr
541-16N &     & V        &  14.19& 0.90& 1.08& 11.79& cl& 0.42& ex&  0.5& ex&    --&   3& -no& miDi& -no-& thD& SA0& *\cr
541-16S&      & V        &  14.10& 0.60& 0.62& 10.19& no& 0.75& ex& -1.8& co&   --&   40& bar& -?Di& -no-&  -?-& SB0& *\cr
       & BGP 41& R       &    (1)& 0.74& 0.81&   (1)& cl& 0.70& ex&  2.0& ex&   (1)&  12& bar& -?Di& -no-& spH& SB0& \cr
       & T 40  & 2 V images&  14.31& 0.75& 0.80& 11.17& st& 0.78& ex&  1.7& co&  0.86&   5& -no& emDi& -no-& spH& SA0& \cr
\noalign{\medskip\hrule\medskip}
}

\vfill\eject\null\vfill\eject
{\bf Notes to Table 5 (Perseus galaxies)}
\bigskip

\noindent {\bf T40}: the identification of Tiff 40 is doubtful. The observed 
galaxy is the nearest one to the Tiff 40 coordinates. A brighter (spiral) 
galaxy is 4 arcmin SE, which is more likely Tiff 40, if we believe Kent \& 
Sargent's (1983) classification (S). 

\noindent {\bf 540-32}: companion of NGC 1177, roundish.
 
\noindent {\bf 540-33}: peanut shaped SA0.
 
\noindent {\bf 540-41}: face on.
 
\noindent {\bf 540-46}: the small field of view did not allow us to image the 
whole envelope. 
 
\noindent {\bf 540-57}: there is a bump on the major axis SuBr profile at V=23 
mag arcsec$^2$. 
 
\noindent {\bf 540-73}: this galaxy is 19 arcsec from a very bright star, 
preventing the detailed structural analysis. However, bulge and disk are 
clearly visually detected as a change of axis ratio and as a bump on the major 
axis profile, leading to our estimated type.  The galaxy is edge-on. 
 
\noindent {\bf 540-74}: our morphological type fully confirms that of Poulain, 
Nieto \& Davoust (1992). 

\noindent {\bf 540-76}: strong, but smooth, dust lane on major axis. A second 
faint dust lane is present with an inclination of about 10 degrees with 
respect to the galaxy major axis 10 arcsec Southward. 
 
\noindent {\bf 540-81}: dust just in the inner part of this galaxy observed 
near the CCD's edge. 
 
\noindent {\bf 540-89}: a very brigh star 20 arcsec from the galaxy center 
perturbs the isophotal analysis at that radius. $e_4$ never dominates the 
Fourier terms, $f_4$ asymmetry. 
 
\noindent {\bf 540-106}: almost edge-on, a bit S-shaped, asymmetric.
 
\noindent {\bf 540-114}: the c/a profile starts to rise at fainter SuBr than 
sampled by plate, thus explaining the difference in the envelope 
classification. 
 
\noindent {\bf 541-15S}: face on.

\noindent {\bf 541-16N}: the galaxy Zw 540-16S and nearby stars prevented us 
from measuring the shape of this galaxy's envelope. 

\noindent {\bf 541-16S}: the galaxy 540-16N makes it impossible to measure the 
SuBr profile at $r>12$ arcsec.

\vfill\eject\null\vfill\eject

\bye